\documentclass{article}

\title{\bf Derivation of Lorentz Invariance and Three Space Dimensions 
in Generic Field Theory}

\author{C~D.~Froggatt\footnote{E-mail:
c.froggatt@physics.gla.ac.uk}\\
{\it Department of Physics and Astronomy,
Glasgow University,}\\
{\it Glasgow G12 8QQ, Scotland}
\and
H.~B.~Nielsen\footnote{E-mail: hbech@mail.desy.de; %
hbech@nbi.dk}\\
{\it Deutsches Elektronen-Synchrotron DESY,
Notkestra{\ss}e 85,}\\
{\it D-22603 Hamburg, Germany}\\
{\it and}\\
{\it The Niels Bohr Institute,
Blegdamsvej 17, Copenhagen {\O}, Denmark}}

\begin{document}
\begin{flushright}
GUTPA/02/10/01\\
\end{flushright}
\vskip .1in
\begin{center}

{\LARGE \bf Derivation of Lorentz Invariance \\
\vspace{6pt}
and Three Space Dimensions \\
\vspace{6pt}
in Generic Field Theory
}

\vspace{20pt}

{\bf \large C.~D.~Froggatt}

\vspace{6pt}

{ \em Department of Physics and Astronomy,\\
 Glasgow University, Glasgow G12 8QQ,
Scotland\\}
\vspace{10pt}

{\bf \large H.~B.~Nielsen}

\vspace{6pt}

{ \em Deutsches Elektronen-Synchrotron DESY,\\
Notkestra{\ss}e 85,
D-22603 Hamburg, Germany}\\
{\it and}\\
{\it The Niels Bohr Institute,
Blegdamsvej 17, Copenhagen {\O}, Denmark}

\end{center}
\vspace{6pt}
\section*{ }
\begin{center}
{\large\bf Abstract}
\end{center}

\date{}

A very general quantum field theory, which is not even
assumed to be Lorentz invariant, is studied in the limit of very
low energy excitations. Fermion and Boson field theories are
considered in parallel. Remarkably, in both cases it is argued
that, in the free and lowest energy approximation, a relativistic
theory with just three space and one time dimension emerges for
each particle type separately. In the case of Fermion fields it is
in the form of the Weyl equation, while in the case of the Bosons
it is essentially in the form of the Maxwell equations.
\vspace{120pt}

To be published in the 
Proceedings of the International Workshop on 
{\it What comes beyond the Standard Model}, Bled, Slovenia, 
July 2002.

\thispagestyle{empty}
\newpage

\section{Introduction}

Since many years ago~\cite{RDold}, we have worked on the project of
``deriving'' all the known laws of nature, especially the symmetry
laws~\cite{book}, from the assumption of the existence of
exceedingly complicated fundamental laws of nature. However the
derivations are such {\em that it practically does not
matter what these exceedingly complicated laws are in detail, just
provided we only study them in some limits } such as the low energy
limit. This is the project which we have baptized ``Random dynamics'',
in order to make explicit the idea that we are thinking of
the fundamental laws of nature as being given by a particular model
pulled out at random from a very large class of
models. In this way, one can overcome the immediate reproach to the
project that it is easy to invent model-proposals which, indeed, do not
deliver the laws of nature as we know them today. We only make the
claim that sufficiently
complicated and generic models should work, not very special ones that
could potentially be constructed so as  not to work.
Also it should be stressed
that there is a lot of interpretation involved,
as to which elements in the ``random'' model
are to be identified phenomenologically with what.
As a consequence, the project tends to
be somewhat phenomenological itself, honestly speaking.
However, in principle,
we should only use the phenomenology to find out which quantities in the
``random dynamics'' model are to be identified with which physically defined
quantities (concepts).

One of the most promising steps, in developing this random dynamics
project, was~\cite{RDold,book} to start without assuming
Lorentz invariance but to assume that we already have several
known laws such as quantum mechanics, quantum
field theory and momentum conservation. Lorentz invariance was then
``derived'', at least for a single species of Weyl particles which
emerged at low energy. However this ``derivation'' of Lorentz invariance
might not actually be the most interesting result from this step in
random dynamics; it is after all not such an overwhelming success,
since it only works for one particle species on its own and does not,
immediately at least, lead to Lorentz invariance if several particle
species are involved. It may rather be the prediction of the number
of space dimensions which is more significant.
Actually the fundamental model is assumed to have an arbitrary number of
dimensions and has momentum degrees of freedom in all these dimensions,
but the velocity components in all but three
dimensions turn out to be zero. In this way the extra dimensions
are supposedly not accessible. So the prediction is effectively
that there are just three spatial dimensions (plus one time)!

In these early studies only a fermionic field theory (without Lorentz
symmetry) was considered, while Bosons were left out of consideration;
we then sometimes speculated that the Bosons could at least be
partly composed from Fermions and thus inherit their Lorentz symmetry.
Indeed, even in more recent work, it is the Fermions that play the main
role~\cite{NormaHolger,RughHolger}. For a summary of other recent 
theoretical models and experimental tests of Lorentz invariance 
breaking see, for example, reference \cite{Kostelecy}.  

It is the purpose of the present paper to review the work with Fermions
stressing a new feature aimed at solving a certain technical
problem---the use of the ``Homolumo-gap-effect'' to be explained
below---and to extend the work to the case of bosonic fields, which is
a highly non-trivial extension.

In the following section we shall put forward our very general
field theory model and then, in section 3, we shall write down in
parallel the equations of motion for Bosons and Fermions
respectively. It turns out that we obtain a common equation of
motion for the ``fields'' in ``momentum''
representation---momentum here being really thought of as a rather
general parameterisation of the degrees of freedom, on which the
Hamiltonian and commutation rules depend smoothly. This equation
of motion involves an antisymmetric matrix which depends on the
``momenta''. The behaviour of the eigenvalue spectrum of such an
antisymmetric real matrix is studied in section 4, with the help
of some arguments based on the Homolumo-gap-effect which are
postponed till section 5. The conclusions are put into section 6.

\section{A random dynamics model}

Since it is our main purpose to derive Lorentz symmetry 
together with $3 + 1$
dimensions, we must start from a model that does not
assume Lorentz invariance nor the precise number of space dimensions
from the outset.
We would, of course, eventually hope to avoid having to assume
momentum conservation or even the existence of the concept of
momentum. However this assumption is less crucial than the others,
since the derivation of Lorentz invariance is highly non-trivial
even if momentum conservation is assumed. Therefore, ``for
pedagogical reasons'', we shall essentially assume translational
symmetry and momentum conservation in our model---in practice
though we shall actually allow a small departure from
translational symmetry. That is to say we consider the model
described in terms of a Fock space, corresponding to having
bosonic or fermionic particles that can be put into single
particle states which are momentum eigenstates. This gives rise to
bosonic and fermionic
fields $\phi(\vec{p})$ and $\psi(\vec{p})$ annihilating these
particles. We shall formulate the model in terms of fields that
are essentially real or obey some Hermiticity conditions, which
mean that we can treat the fields $\phi(\vec{p})$ and
$\psi(\vec{p})$ as Hermitian fields. In any case, one can always
split up a non-Hermitian field into its Hermitian and
anti-Hermitian parts. This is done since, in the spirit of the
random dynamics project, we do not want to assume any charge
conservation law from the outset.

\subsection{Technicalities in a general momentum description}

In the very general type of model we want to set up, without any
assumed charge conservation, it is natural to
use a formalism which is suitable for neutral particles like,
say, $\pi^0$ mesons.  However,
when one constructs a second quantized formalism
from a single particle Fock-space description,
in which there can be different numbers of particles in the different single
particle states\footnote{In the fermionic case there can be 0 or 1
particle in a particular single particle state, while in the bosonic
case there can also be many.}, one at first gets
{\em ``complex''} i.e.~{\em non-Hermitian}
second quantized fields. In order to describe say the $\pi^0$-field,
one must put
restrictions on the allowed Fock-space states, so that one cannot
just completely freely choose how many particles there should be in each
single particle state. Basically one ``identifies'' particles and
antiparticles (= holes), so that they are supposed to be in analogous
states (in the Fermion case, it is the Majorana condition that must be
arranged). Field creation of a particle with momentum $\vec{p}$ is
brought into correspondence with annihilation of a particle
with momentum $-\vec{p}$.

In our general description of bosonic or fermionic second
quantized particles, we want to use a formalism of this $\pi^0$ or
Majorana type. We can always return to
a charged particle description by introducing a doubling of the
number of components for such a field; we can simply make a
non-Hermitian (i.e.~essentially charged) field component from two
Hermitian ones, namely the Hermitian (``real'') and
anti-Hermitian (``purely imaginary'') parts, each of which are
then Majorana or $\pi^0$-like. Let us recall here that the $\pi^0$
field is Hermitian when written as a field depending on the position
variable $\vec{x}$, while it is not Hermitian in momentum space.
In fact, after Fourier transformation, the property of
Hermiticity or reality in position space becomes the property,
in momentum representation, that the fields at $\vec{p}$
and $-\vec{p}$ are related by
Hermitian/complex conjugation:
\begin{equation}
\label{reflection}
\phi(\vec{p}) = \phi^{\dagger}(-\vec{p})
\end{equation}

For generality, we should also like to have
Hermitian momentum dependent fields, which corresponds to having
a similar reflection symmetry in position space, saying that
the values of the fields at $\vec{x}$ and $-\vec{x}$
are related by Hermitian/complex conjugation.
To make the ``most general'' formalism for our study, we should
therefore impose Hermiticity both in momentum {\em and} in
position representation. We then
have to accept that we also have a reflection symmetry in both
position and momentum space. In this paper, we shall
in reality only
consider this most general formalism for bosonic fields.
For this purpose, let us denote the $\pi^0$ field and its
momentum conjugate field by $\phi_0(\vec{p})$ and $\pi_0(\vec{p})$
respectively. Then, in standard relativistic quantum field theory,
the non-vanishing equal time commutation
relations between their real and imaginary parts are as
follows:
\begin{eqnarray}
\label{pi0commutators}
\left[ \mathrm{Re}\,\phi_0(\vec{p}),
\mathrm{Re}\,\pi_0(\vec{p'}) \right]
=\frac{i}{2} \left(\delta (\vec{p} - \vec{p'})
+\delta(\vec{p} + \vec{p'}) \right) \\
\left[ \mathrm{Im}\,\phi_0(\vec{p}),
\mathrm{Im}\,\pi_0(\vec{p'}) \right]
=\frac{i}{2} \left(\delta (\vec{p} - \vec{p'})
-\delta(\vec{p} + \vec{p'}) \right)
\end{eqnarray}
We note that the appearance of the $\delta(\vec{p} + \vec{p'})$
function as well as the $\delta(\vec{p} - \vec{p'})$ function is a
consequence of the reflection symmetry (\ref{reflection}).

Now the reader should also notice that we are taking the point of
view that many of the observed laws of nature are only laws of
nature in the limit of {\em ``the poor physicist''}, who is
restricted to work with the lowest energies and only with a small
range of momenta compared to the fundamental (Planck) scale. In
the very generic and not rotational invariant type of model which
we want to consider, it will now typically happen that the small
range of momenta to which the physicist has access is not centred
around zero momentum---in the presumably rather arbitrary choice
of the origin for momentum---but rather around some momentum,
$\vec{p}_0$ say. This momentum $\vec{p_0}$ will generically be
large compared to the momentum range accessible to the poor
physicist; so the reflection symmetry in momentum space and the
associated $\delta(\vec{p} + \vec{p'})$ terms in commutators will
not be relevant to the poor physicist and can be ignored. However,
in our general field theory model, there can be a remnant
reflection symmetry in position space. Indeed we shall see below
that what may be considered to be a mild case of momentum
non-conservation does occur for the Maxwell equations derived in
our model: there is the occurrence of a reflection centre
somewhere, around which the Maxwell fields should show a parity
symmetry in the state of the fields. If we know, say, the electric
field in some place, then we should be able to conclude from this
symmetry what the electric field is at the mirror point. If, as is
most likely, this reflection point is far out in space, it would
be an astronomical challenge to see any effect of this lack of
translational symmetry. In this sense the breaking of
translational symmetry is very ``mild''.

\subsection{General Field Theory Model}
At the present stage in the development of our work, it is assumed
that we {\em only work to the free field approximation} and thus
the Hamiltonian is taken to be bilinear in the Hermitian fields
$\psi(\vec{p})$ and $\phi(\vec{p})$. Also, because of the assumed
rudiment of momentum conservation in our model, we only consider
products of fields taken for the same momentum $\vec{p}$.
In other words our Hamiltonian takes the following form:
\begin{equation}
H_F =\frac{1}{2} \int d\vec{p} \:
\sum_{i,j}\psi_i(\vec{p})\psi_j(\vec{p})
H_{ij}^{(F)}(\vec{p})
\end{equation}
and
\begin{equation}
H_B =\frac{1}{2} \int d\vec{p} \:
\sum_{i,j}\phi_i(\vec{p})
\phi_j(\vec{p})H_{ij}^{(B)}(\vec{p}).
\end{equation}
for Fermions and Bosons respectively.
Here the coefficient functions $H_{ij}^{(F)}(\vec{p})$ and
$H_{ij}^{(F)}(\vec{p})$ are non-dynamical in the free field
approximation and just reflect the general features of ``random''
laws of nature expected in the random dynamics project. That is to
say we do not impose Lorentz invariance conditions on these
coefficient functions, since that is what is hoped to {\em come
out} of the model. We should also not assume that the $\vec{p}$
vectors have any sort of Lorentz transformation properties {\it a
priori} and they should not even be assumed to have, for instance,
3 spatial dimensions. Rather we start out with $D>3$
spatial dimensions;
then one of our main achievements will be to show that the
velocity components in all but a three dimensional subspace are
zero. It is obvious that, in these expressions, the coefficient
functions  $H_{ij}^{(F)}(\vec{p})$ and $H_{ij}^{(F)}(\vec{p})$ can
be taken to have the symmetry properties:
\begin{equation}
 H_{ij}^{(F)}(\vec{p})=-  H_{ji}^{(F)}(\vec{p})\quad \quad
\hbox{ and}\quad\quad
H_{ij}^{(B)}(\vec{p})=
H_{ji}^{(B)}(\vec{p}).
\end{equation}

However, it should be borne in mind that {\it a priori} the fields are
arbitrarily normalised and that we may use the Hamiltonians
to define the normalisation of the fields, if we so choose.
In fact an important ingredient in the formulation of the present
work is to assume that a linear transformation has been made on the
various field components $\phi_i(\vec{p})$, i.e.~a transformation
on the component index $i$, such that the symmetric coefficient
functions $H_{ij}^{(B)}(\vec{p})$ become equal to the unit matrix:
\begin{equation}
H_{ij}^{(B)}(\vec{p}) = \delta_{ij} \qquad
\hbox{(by normalisation for all $\vec{p}$)}
\end{equation}
Thereby, of course, the commutation relations
among these components $\phi_i(\vec{p})$ are modified and
we cannot simultaneously arrange for them to be trivial. So for the Bosons
we choose a notation in which the non-trivial behaviour of
the equations of motion, as a function of  the momentum
$\vec{p}$, is put into
the commutator expression\footnote{Note that we are here
ignoring possible terms of the form
$iB_{ij}(\vec{p}) \delta(\vec{p} + \vec{p'})$ as irrelevant
to the poor physicist, according to the discussion after equation
(\ref{pi0commutators}).}
\begin{equation}
[\phi_i(\vec{p}), \phi_j(\vec{p'})]=iA_{ij}(\vec{p}) \delta(\vec{p} - \vec{p'})
\label{commutator}
\end{equation}
It follows that the information which we would, at first,
imagine should be contained in the Hamiltonian is, in fact,
now contained in the antisymmetric matrices $A_{ij}(\vec{p})$.

For the Fermions, on the other hand, we shall keep to the more
usual formulation. So we normalize the anti-commutator to be the
unit matrix and let the more
nontrivial dependence on $\vec{p}$ sit in the Hamiltonian
coefficient functions $H_{ij}^{(F)}(\vec{p})$. That is to say that
we have the usual equal time anti-commutation relations:
\begin{equation}
\{ \psi_i(\vec{p}), \psi_j(\vec{p'})\} =
\delta_{ij} \delta(\vec{p} - \vec{p'}).
\end{equation}

The component indices $i$, $j$ enumerate the very general discrete
degrees of freedom in the model. These degrees of freedom might, at the
end, be identified with Hermitian and anti-Hermitian components, spin
components, variables versus conjugate momenta or even totally
different types of particle species, such as flavours etc.
It is important to realize that this model
is so general that it has, in that sense, almost no assumptions built
into it---except for our free approximation, the above-mentioned rudimentary
momentum conservation and some general features of second quantized models.
It follows from the rudimentary momentum conservation in our model that the
(anti-)commutation relations have a $\delta(\vec{p} - \vec{p'})$ delta
function factor in them.

Obviously the Hermiticity of the Hamiltonians for the second quantized
systems means that the matrices $H_{ij}^{(F)}(\vec{p})$ and
$H_{ij}^{(B)}(\vec{p})$ are Hermitian
and thus have purely imaginary and real matrix elements respectively.
Similarly, after the extraction of the $i$ as a conventional factor in
equation (\ref{commutator}), the matrix $A_{ij}(\vec{p})$
has real matrix elements and is antisymmetric.

\section{Equations of motion for the general fields}
\vspace{.3cm}

\noindent
We can easily write down the equations of motion for the field components
in our general quantum field theory, both in the fermionic case:
\begin{equation}
\dot{\psi}_i(\vec{p}) = i[H_F, \psi_i(\vec{p})] =
 i \sum_k\psi_k(\vec{p})H^{(F)}_{ki}(\vec{p})
\end{equation}
and in the bosonic case:
\begin{equation}
\label{eq8} \dot{\phi}_i(\vec{p}) = i[H_B, \phi_i(\vec{p})] = -
\sum_k\phi_k(\vec{p})A_{ki}(\vec{p}).
\end{equation}

Since $H^{(F)}_{ij}(\vec{p})$ has purely imaginary matrix elements,
we see that both the bosonic and the fermionic equations of motion
are of the form
\begin{equation}
\dot{\xi_i}(\vec{p})= \sum_k A_{ik}\xi_k(\vec{p})
\end{equation}
In the fermionic case we have extracted a factor of $i$,
by making the definition
\begin{equation}
H^{(F)}_{ij}(\vec{p}) = i A_{ij}(\vec{p}).
\end{equation}
Also the Boson field $\phi$ and the Fermion field $\psi$ have both
been given the neutral name $\xi$ here.

\section{Spectrum of an antisymmetric matrix depending on $\vec{p}$}

An antisymmetric matrix $A_{ij}(\vec{p})$ with real matrix elements is
anti-Hermitian and thus has purely imaginary eigenvalues. However, if
we look for a time dependence ansatz of the form
\begin{equation}
\xi_i(\vec{p},t)= a_i(\vec{p}) \exp(-i\omega t),
\end{equation}
the eigenvalue equation for the frequency $\omega$ becomes
\begin{equation}
\omega a_i= \sum_j iA_{ij}(\vec{p})a_j.
\end{equation}
Now the matrix $iA_{ij}(\vec{p})$ is Hermitian and the
eigenvalues $\omega$ are therefore real.

It is easy to see, that if $\omega$ is an eigenvalue, then so also is
$-\omega$. In fact we could imagine calculating the eigenvalues by
solving the equation
\begin{equation}
\det{(iA - \omega)} =0
\end{equation}
We then remark that transposition of the matrix $(iA-\omega)$ under the
determinant sign will not change the value of the determinant, but
corresponds to changing the sign of $\omega$ because of the antisymmetry
of the matrix $iA$.
So non-vanishing eigenvalues occur in pairs.

In order to compare with the more usual formalism, we should
really keep in mind that the creation operator for a particle with
a certain $\omega$-eigenvalue is, in fact, the annihilation
operator for a particle in the eigenstate with the opposite value
of the eigenvalue, i.e.~$-\omega$. Thus, when thinking in usual
terms, we can ignore the negative $\omega$ orbits as being already
taken care of via their positive $\omega$ partners. The unpaired
eigenstate, which is formally a possibility for $\omega=0$, cannot
really be realized without some little ``swindle''. In the bosonic
case it would correspond to a degree of freedom having, say, a
generalized coordinate but missing the conjugate momentum. In the
fermionic case, it would be analogous to the construction of a set
of $\gamma$-matrices in an odd dimension, which is strictly
speaking only possible because one allows a relation between them
(the product of all the odd number of them being, say, unity) or
because one allows superfluous degrees of freedom.
It is obviously difficult to construct such a set of
$\gamma$-matrices in complete analogy with the case of
an even number of fields, since then the
number of components in the representation of the $n$
gamma-matrices would be $2^{n/2}$, which can hardly make sense for
$n$ odd. Nevertheless, we shall consider the possibility of an
unpaired $\omega=0$ eigenstate in the bosonic case below.

Now the main point of interest for our study is how the second quantized
model looks close to its
ground state.  The neighbourhood of this ground state is supposed to
be the only regime which we humans can study in our ``low energy''
experiments, with small momenta compared to the fundamental (say
Planck) mass scale.
With respect to the ground state of such a second quantized world machinery,
it is well-known that there is a difference between the fermionic and the
bosonic case. In the fermionic case, you can at most have one Fermion in
each state and must fill the states up to some specific value of the
single particle energy---which is really $\omega$. However, in the bosonic
case, one can put a large number of Bosons into the same orbit/single
particle state, if that should pay energetically.

\subsection{ The vacuum}
If we allow for the existence of a chemical potential, which
essentially corresponds to the conservation of the number of
Fermions, we shall typically get the ground state to have Fermions
filling the single particle states up to some special value of the
energy called the Fermi-energy $\omega_{FS}$ ($FS$ standing for
``Fermi-surface''). For Bosons, on the other hand, we will always
have zero particles in all the orbits, except perhaps in the zero
energy ground state; it will namely never pay energetically to put
any bosons into positive energy orbits.

\subsection{The lowest excitations}

So for the investigation of the lowest excitations, i.e.~those that a
``poor physicist'' could afford to work with, we should look for
the excitations very near to the Fermi-surface in the fermionic
case. In other
words, we should put Fermions into the orbits with energies very little
above the Fermi-energy,
or make holes in the Fermi-sea at values of the orbit-energies very
little below the Fermi-energy. Thus, for excitations accessible to
the ``poor physicist'', it is only necessary to study the behaviour
of the spectrum for the Bosons having a value of $\omega$ near to zero,
and for the Fermions having a value of $\omega$ near the Fermi-energy
$\omega_{FS}$.

\subsubsection{Boson case: levels approaching a group of
$\omega =0$ levels}
In section 5 we shall argue that, if the model
has adjustable degrees of freedom (``garbage variables''), they
would tend to make the $\omega = 0$ eigenvalue multiply
degenerate. However, for simplicity, we shall first consider here
the case where there {\em is} just a single zero-eigenvalue
$\omega$-level. We should mention that the true generic situation
for an even number of fields is that there are normally no
zero-eigenvalues at all. So what we shall study here, as the
representative case, really corresponds to the case with an odd
number of fields. In this case there will normally be just one
(i.e.~non-degenerate) $\omega=0$ eigenvalue. However it can happen
that, for special values of the ``momentum parameters'', a pair of
eigenvalues---consisting of eigenvalues of opposite sign of
course---approach zero. It is this situation which we believe to
be the one of relevance for the low energy excitations.

We shall concentrate our interest on a small region in the
momentum parameter space, around a point $\vec{p}_0$
where the two levels with
the numerically smallest non-zero eigenvalues merge together with
a level having zero eigenvalue. Using the well-known fact that, in
quantum mechanics, perturbation corrections from faraway levels
have very little influence on the perturbation of a certain level,
we can ignore all the levels except the zero eigenvalue and this
lowest non-zero pair. So if, for simplicity, we think of this case
of just one zero eigenvalue except where it merges with the other
pair, we need only consider three states and that means, for the
main behaviour, we can calculate as if there were only the three
corresponding fields. This, in turn, means that we can treat the
bosonic model in the region of interest, by studying the spectrum
of a (generic) antisymmetric $3\times 3$ matrix with real
elements, or rather such a matrix multiplied by $i$. Let us
immediately notice that such a matrix is parameterised by
\underline{three} parameters. The matrix and thus the spectrum, to
the accuracy we are after, can only depend on three of the
momentum parameters. In other words the dispersion relation will
depend trivially on all but 3 parameters in the linear
approximation. By this linear approximation, we here mean the
approximation in which the ``poor physicist'' can only work with
a small region in momentum parameter space also---not only in
energy. In this region we can trust the lowest order Taylor
expansion in the differences of the momentum parameters from their
starting values (where the nearest levels merge). Then the
$\omega$-eigenvalues---i.e.~the dispersion relation---will not
vary in the direction of a certain subspace of co-dimension three.
Corresponding to these directions the velocity components of the
described Boson particle will therefore be zero! The Boson, as
seen by the ``poor physicist'', can only move inside a three
dimensional space; in other directions its velocity must remain
zero. It is in this sense we say that the three-dimensionality of
space is explained!

\subsection{Maxwell equations}
\label{Maxwelleq}

The form of the equations of motion for the fields, in this low excitation
regime where one can use the lowest order Taylor expansion in the momentum
parameters, is also quite remarkable: after a linear transformation in
the space of ``momentum parameters'', they can be transformed into the
{\em Maxwell equations} with the fields being complex (linear)
combinations of the magnetic and electric fields.

We can now easily identify the linear combinations of the momentum
parameters minus their values at the selected merging point, which
should be interpreted as true physical momentum components. They
are, in fact, just those linear combinations which occur as matrix
elements in the $3\times 3$ matrix $A$ describing the development
of the three fields $\phi_j$ relevant to the ``poor physicist''.
That is to say we can choose the definition of the ``true momentum
components'' $\vec{k}$ as such linear functions of the deviations,
$\vec{p} -\vec{p}_0$, of the momentum parameters from the merging
point that the antisymmetric matrix $A$ reduces to
\begin{equation}
\label{eq14}
A = \left ( \begin{array}{c|c|c}0 & k_3 & -k_2\\
-k_3 & 0 & k_1\\ k_2 & -k_1 & 0\\ \end{array} \right)
\end{equation}
with eigenvalues $-i\omega = 0,\pm i\sqrt{k_1^2+k_2^2+k_3^2}$.

In the here chosen basis for the momenta, we can make a Fourier
transform of the three fields $\phi_j(\vec{k})$ into the
$\vec{x}$-representation.
These new position space fields $\phi_j(\vec{x})$ are no
longer Hermitian. However,
it follows from the assumed Hermiticity
of the $\phi_j(\vec{k})$ that, in the $\vec{x}$-representation,
the real parts of the fields $\phi_j(\vec{x})$ are even, while the
imaginary parts are odd functions of $\vec{x}$.
We now want to identify these
real and imaginary parts as magnetic and electric fields
$B_j(\vec{x})$ and $E_j(\vec{x})$ respectively: $\phi_j(\vec{x}) =
iE_j(\vec{x}) + B_j(\vec{x})$. However the symmetry of these
Maxwell fields means that they must be in a configuration/state
which goes into itself under a parity reflection in the origin.
This is a somewhat strange feature which seems necessary for the
identification of our general fields with the Maxwell fields; a
feature that deserves further investigation. For the moment let
us, however, see that we do indeed get the Maxwell equations in
the free approximation with the proposed identification.

By making the inverse Fourier transformation back to momentum
space, we obtain the following identification of the fields
$\phi_j(\vec{k})$ in our general quantum field theory with the
electric field $E_j(\vec{k})$ and magnetic field $B_j(\vec{k})$
Fourier transformed into momentum space:
\begin{equation}
\label{identification}
\left ( \begin{array}{c}\phi_1(\vec{k})\\
\phi_2(\vec{k})\\\phi_3(\vec{k})\\ \end{array} \right)=
\left ( \begin{array}{c}iE_1(\vec{k}) +  B_1(\vec{k})\\
iE_2(\vec{k}) + B_2(\vec{k})\\i E_3(\vec{k}) +B_3(\vec{k})\\
\end{array} \right).
\end{equation}
We note that the Fourier transformed electric field $E_j(\vec{k})$
in the above ansatz (\ref{identification}) has to be purely
imaginary, while the magnetic field $B_j(\vec{k})$ must be purely
real.

By using the above identifications, eqs.~(\ref{eq14}) and
(\ref{identification}), the equations of motion (\ref{eq8}) take
the following form
\begin{equation}
\left ( \begin{array}{c}i\dot{E}_1(\vec{k}) +  \dot{B}_1(\vec{k})\\
i\dot{E}_2(\vec{k}) + \dot{B}_2(\vec{k})\\i \dot{E}_3(\vec{k})
+ \dot{B}_3(\vec{k})\\
\end{array}
\right)  =\left ( \begin{array}{c|c|c}0 & k_3 & -k_2\\
-k_3 & 0 & k_1\\ k_2 & -k_1 & 0\\ \end{array} \right)
\left ( \begin{array}{c}iE_1(\vec{k}) +  B_1(\vec{k})\\
iE_2(\vec{k}) + B_2(\vec{k})\\i E_3(\vec{k}) +B_3(\vec{k})\\
\end{array} \right).
\end{equation}
We can now use the usual Fourier transformation identification in
quantum mechanics to transform these equations
to the $\vec{x}$-representation,
simply from the definition of $\vec{x}$ as the Fourier transformed
variable set associated with $\vec{k}$,
\begin{equation}
k_j  = i^{-1} \partial_j
\end{equation}
Thus in $\vec{x}$-representation the equations of motion become
\begin{equation}
\left ( \begin{array}{c}i\dot{E}_1(\vec{x}) +  \dot{B}_1(\vec{x})\\
i\dot{E}_2(\vec{x}) + \dot{B}_2(\vec{x})\\i \dot{E}_3(\vec{x})
+\dot{B}_3(\vec{x})\\
\end{array}
\right) =\left ( \begin{array}{c|c|c}0 & -i\partial_3 & i\partial_2\\
i\partial_3 & 0 & -i\partial_1\\ -i\partial_2 & i\partial_1 & 0\\
\end{array}
\right)\left ( \begin{array}{c}iE_1(\vec{x}) +  B_1(\vec{x})\\
iE_2(\vec{x}) + B_2(\vec{x})\\i E_3(\vec{x}) +B_3(\vec{x})\\
\end{array} \right).
\end{equation}
The imaginary terms in the above equations give rise to the
equation:
\begin{equation}
\label{iprop} \dot{\vec{E}}(\vec{x}) = \mathrm{curl}\, \vec{B}
\end{equation}
while the real parts give the equation:
\begin{equation}
\label{realpart} \dot{\vec{B}}(\vec{x}) =- \mathrm{curl}\, \vec{E}
\end{equation}
These two equations are just the Maxwell equations in the absence
of charges and currents, except that strictly speaking we miss two
of the Maxwell equations, namely
\begin{equation}
\label{missing} \mathrm{div}\, \vec{E}(\vec{x}) =0 \quad
\hbox{ {and} } \quad \mathrm{div}\, \vec{B}(\vec{x}) =0.
\end{equation}
However, these two missing equations are derivable from the other
Maxwell equations in time differentiated form. That is to say, by
using the result that the divergence of a curl is zero, one can
derive from the other equations that
\begin{equation}
\label{timederived}
\mathrm{div}\, \dot{\vec{E}}(\vec{x}) =0 \quad \hbox{
{and}} \quad \mathrm{div}\, \dot{\vec{B}}(\vec{x}) =0
\end{equation}
which is though not quite sufficient. Integration of the resulting
equations (\ref{timederived}) effectively replaces the $0$'s on
the right hand sides of equations (\ref{missing}) by terms
constant in time, which we might interpret as some constant
electric and magnetic charge distributions respectively. In our
free field theory approximation, we have potentially ignored such
terms. So we may claim that, in the approximation to which we have
worked so far, we have derived the Maxwell equations sufficiently
well.

\section{Homolumo-gap and analogue for bosons}

The Homolumo-gap effect refers to a very general feature of
systems of Fermions, which possess some degrees of freedom that
can adjust themselves so as to lower the energy as much as
possible. The effect is so general that it should be useful for
almost all systems of Fermions, because even if they did not have
any extra degrees of freedom to adjust there would, in the Hartree
approximation, be the possibility that the Fermions could
effectively adjust themselves. The name Homolumo gap was
introduced in chemistry and stands for the gap between `` the
highest occupied'' HO ``molecular orbit'' MO and the ``lowest
unoccupied'' LU ``molecular orbit'' MO.  The point is simply that
if the filled (occupied) orbits (single particle states) are
lowered the whole energy is lowered, while it does not help to
lower the empty orbits. It therefore pays energetically to make
the occupied orbits go down in energy and separate from the unfilled
ones; thus a gap may appear or rather {\em there will be a general
tendency to get a low level density near the Fermi-surface}. This
effect can easily be so strong that it causes a symmetry to break
\cite{Teller}; symmetry breaking occurs if some levels, which are
degenerate due to the symmetry, are only partially filled so that
the Fermi-surface just cuts a set of degenerate states/orbits. It
is also the Homolumo-gap effect which causes the deformation of
transitional nuclei, which are far from closed shell
configurations. We want to appeal to this Homolumo gap
effect, in subsection \ref{Weylderivation}, as a justification for
the assumption that the Fermi-surface gets close to those places
on the energy axis where the level density is minimal.

However we first want to discuss a similar effect, where the degrees
of freedom of a system of Bosons adjust themselves to lower the
total energy. As for the Fermion systems just discussed, this
lowering of the total energy is due to the adjustment of a sum
over single particle energies---the minimisation of the zero-point
energy of the bosonic system. We consider the effect of this
minimisation to be the analogue for Bosons of the Homolumo-gap
effect.

\subsection{The analogue for bosons}
In the ``derivation'' of the Maxwell equations given in subsection
\ref{Maxwelleq}, we started by introducing the assumption of the
existence of a zero frequency, $\omega=0$, eigenvalue by taking
the number of Hermitian fields and thereby the order of the
antisymmetric matrix $A_{ij}$ to be {\em odd.} We now turn to our
more general assumption of the existence of multiply degenerate
$\omega=0$ eigenvalues. Honestly we can only offer a rather
speculative argument in favour of our assumption that there should
be several eigenvalues which are zero, even in the case when the
total number of fields is not odd. For quite generic matrices, as
would be the cleanest philosophy, it is simply not true that there
would be zero eigenvalues for most momenta in the case of an even
number of fields. However, let us imagine that there are many
degrees of freedom of the whole world machinery that could adjust
themselves to minimize the energy of the system and could also
influence the matrix $A_{ij}(\vec{p})$. Then one could, for
instance, ask how it would be energetically profitable to adjust
the eigenvalues, in order to minimize the zero-point energy of the
whole (second quantized) system. This zero-point energy is
formally given by the integral over all (the more than three
dimensional) momentum space; let us just denote this integration
measure by $d\vec{p}$, so that:
\begin{equation}
\label{zpe}
E_{{\it zero-point}} = \int  d\vec{p}
\sum_{{\it eigenvalue\,\, pair   \,\, k}}
|\omega_k(\vec{p})|/2 
\end{equation}
Provided some adjustment took place in minimizing this quantity,
there would {\em a priori} be an argument in favour of having
several zero eigenvalues, since they would contribute the least to
this zero-point energy $E_{{\it zero-point}}$. At first sight,
this argument is not very strong, since it just favours making the
eigenvalues small and not necessarily making any one of them
exactly zero. However, we underlined an important point in favour
of the occurrence of exactly zero eigenvalues, by putting the
numerical sign explicitly into the integrand
$|\omega_k(\vec{p})|/2$ in the expression (\ref{zpe}) for the
zero-point energy. The important point is that the numerical value
function is not an ordinary analytic function, but rather has a
kink at $\omega_k(\vec{p})=0$. This means that, if other
contributions to the energy of the whole system are
smooth/analytic, it could happen that the energy is lowered when
$\omega_k(\vec{p})$ is lowered numerically for both signs of
$\omega_k(\vec{p})$; here we consider $\omega_k(\vec{p})$ to be a
smooth function of the adjusting parameters of the whole world
machinery (we could call them ``garbage parameters''). For a
normal analytic energy function this phenomenon could of course
never occur, except if the derivative just happened (is fine-tuned
one could say) to be equal to zero at $\omega_k(\vec{p}) =0$. But
with a contribution that has the numerical value singularity
behaviour it is possible to occur with a finite probability
(i.e.~without fine-tuning), because it is sufficient that the
derivative of the contribution to the total energy from other
terms
is numerically lower than the derivative of the zero-point term
discussed. Then, namely, the latter will dominate the sign of the
slope and the minimum will occur {\em exactly} for zero
$\omega_k(\vec{p})$.

In this way, we claim to justify our assumption that the matrix
$A_{ij}(\vec{p})$ will have several exactly zero eigenvalues and
thus a far from maximal rank; the rank being at least piecewise
constant over momentum space. We shall therefore now study
antisymmetric matrices with this property in general and look for
their lowest energy excitations.

\subsection{Using several zero eigenvalues to derive Maxwell equations}

As in subsection \ref{Maxwelleq}, we assume that when a single
pair of opposite sign eigenvalues approach zero as a function of the
momentum, we can ignore the faraway eigenvalues. Then, using the
approximation of only considering the fields corresponding to the
two eigenvalues approaching zero and the several exact zero
eigenvalues, we end up with an effective $(n+2)$ x $(n+2)$ matrix
$A_{ij}(\vec{p})$ obeying the constraint of being of rank two (at
most). Now we imagine that we linearize the momentum dependence of
$A_{ij}(\vec{p})$ on $\vec{p}$ around a point in momentum space,
say $\vec{p}_0$, where the pair of eigenvalues approaching zero
actually reach zero, so that the matrix is totally zero,
$A_{ij}(\vec{p}_0)=0$, at the starting point for the Taylor
expansion. That is to say that, corresponding to different basis
vectors in momentum space, we get contributions to the matrix
$A_{ij}(\vec{p})$ linear in the momentum difference
$\vec{p}-\vec{p}_0$. Now any non-zero antisymmetric matrix is
necessarily of rank at least 2. So the contribution from the first
chosen basis vector in momentum space will already give a matrix
$A_{ij}$ of rank 2 and contributions from other momentum
components should not increase the rank beyond this. A single
basis vector for a set of linearly parameterised antisymmetric
real matrices can be transformed to just having elements (1,2) 
and (2,1) nonzero and the rest zero. In order to avoid a further
increase in the rank of the matrix by adding other linear
contributions, these further contributions must clearly not
contribute anything to matrix elements having both column and row
index different from $1$ and $2$. However this is not sufficient
to guarantee that the rank remains equal to 2. This is easily
seen, because we can construct 4 x 4 antisymmetric matrices, which
are of the form of having 0's on all places (i,j) with both i and
j different from 1 and 2 and have nonzero determinant.

So let us consider 4 by 4 sub-determinants of the matrix $A_{ij}$
already argued to be of the form
\begin{equation}
\label{mform}
\left ( \begin{array}{c|c|c|c|c}0 & A_{12} & A_{13} &\cdots & A_{1n}\\
-A_{12} & 0 & A_{23}& \cdots & A_{2n}\\ -A_{13} & -A_{23} & 0& \cdots &0\\
: & : & 0 & \cdots &0\\
-A_{1n} & -A_{2n} & 0 & \cdots & 0 \end{array} \right).
\end{equation}
Especially let us consider a four by four sub-determinant along the
diagonal involving columns and rows 1 and 2. The determinant is
for instance
\begin{equation}
\label{sub4}
det
\left ( \begin{array}{c|c|c|c}0 & A_{12} & A_{13} & A_{15}\\
-A_{12} & 0 & A_{23}& A_{25}\\ -A_{13} & -A_{23} & 0&0\\
-A_{15} & -A_{25} & 0 & 0 \end{array} \right ) =-\left( det \left (
\begin{array}{c|c} A_{13} & A_{15}\\
 A_{23}& A_{25} \end{array} \right ) \right )^2.
\end{equation}
In order that the matrix $A_{ij}$ be of rank 2, this determinant
must vanish and so we require that the 2 by 2 sub-matrix
\begin{equation}
\left (
\begin{array}{c|c} A_{13} & A_{15}\\
 A_{23}& A_{25}\\ \end{array} \right)
\end{equation}
must be degenerate, i.e.~of rank 1 only. This means that the
two columns in it are proportional, one to the other. By
considering successively several such selected four by four
sub-matrices, we can easily deduce that all the two columns
\begin{equation}
\left ( \begin{array}{c} A_{13}\\ A_{23}  \end{array} \right),
\left ( \begin{array}{c} A_{14}\\ A_{24}  \end{array} \right), \cdots
\left ( \begin{array}{c} A_{1n}\\ A_{2n}  \end{array} \right)
\end{equation}
are proportional. This in turn means that we can transform them
all to zero, except for say
\begin{equation}
\left ( \begin{array}{c} A_{13}\\ A_{23} \end{array} \right),
\end{equation}
by going into a new basis for the fields
$\phi_k(\vec{p}-\vec{p_0})$. So, finally, we have
transformed the formulation of the fields in such a way that only
the upper left three by three corner of the $A$ matrix is
non-zero. But this is exactly the form for which we argued in
subsection \ref{Maxwelleq} and which was shown to be interpretable
as the Maxwell equations, and moreover the Maxwell equations for
just three spatial dimensions!

\subsection{The Weyl equation derivation \label{Weylderivation}}

Let us now turn to the application of the Homolumo-gap effect to a
system of Fermions in our general field theory model. We shall
assume that the Homolumo-gap effect turns out to be strong enough
to ensure that the Fermi-surface just gets put to a place where
the density of levels is very low. Actually it is very realistic
that a gap should develop in a field theory with continuum
variables $\vec{p}$ labeling the single particle states. That is
namely what one actually sees in an insulator; there is an
appreciable gap between the last filled {\em band} and the first
empty band. However, if the model were totally of this insulating 
type, the poor physicist would not ``see'' anything, because he is 
supposed to be unable to afford to raise a particle from the filled
band to the empty one. So he can only see something if there are
at least {\em some} Fermion single particle states with energy
close to the Fermi-surface.

We shall now divide up our discussion of what happens near the
Fermi-surface according to the number of components of the Fermion
field that are relevant in this neighborhood.
Let us denote by $n$ the number
of Fermion field components, which contribute significantly to
the eigenstates near the Fermi-surface in the small region
of momentum space we choose to consider.

The eigenvalues $\pm\omega$ of $iA_{ij}$ -- which come in
pairs -- correspond to  eigenstates with complex components. Thus
it is really easiest in the fermionic case to ``go back'' to
a complex field notation, by constructing complex fields out of
twice as big a number of real ones. So now we consider the
level-density near the Fermi-surface for $n$ complex Fermion field
components.

\subsection{The case of $n=0$ relevant levels near Fermi-surface}

The $n=0$ case must, of course, mean that there are no levels at
all near the Fermi-surface in the small momentum range considered.
This corresponds to the already mentioned insulator case. The 
poor physicist sees nothing from such regions in momentum space 
and he will not care for such regions at all. Nonetheless
this is the generic situation close to the Fermi surface and
will apply for most of the momentum space.

\subsection{The case of $n$ = 1 single relevant level near the Fermi-surface}

In this case the generic situation will be that, as a certain
component of the momentum is varied, the level will vary
continuously in energy. This is the kind of behaviour observed
in a metal. So there will be a rather smooth density
of levels and such a situation is not favoured by the Homolumo gap
effect, if there is any way to avoid it.

\subsection{The case of $n$ = 2 relevant levels near the Fermi-surface}

In this situation a small but almost trivial calculation is
needed. We must estimate how a Hamiltonian,
described effectively as a 2 by 2 Hermitian matrix $H$ with
matrix elements depending on the momentum $\vec{p}$,
comes to look in the generic case---\i.e. when nothing is
fine-tuned---and especially how the level density behaves. That
is, however, quite easily seen, when one remembers that the three
Pauli matrices and the unit 2 by 2 matrix together form a basis
for the four dimensional space of two by two matrices. All
possible Hermitian 2 by 2 matrices can be expressed as linear
combinations of the three Pauli matrices $\sigma^j$ and the
unit 2 by 2 matrix $\sigma^0$ with real coefficients.
We now consider a linearized Taylor
expansion\footnote{A related discussion of the redefinition 
of spinors has been given in the context of the low 
energy limit of a Lorentz violating QED model \cite{Colladay}.} 
of the momentum dependence of such matrices, by taking
the four coefficients to these four matrices to be arbitrary
linear functions of the momentum minus the ``starting momentum''
$\vec{p_0}$, where the two levels become degenerate with 
energy $\omega(\vec{p_0})$.
That is to say we must take the Hermitian 2 by 2
matrix to be
\begin{equation}
H = \sigma^a V^i_a (p_i - p_{0i}) + 
\sigma^0 \omega(\vec{p_0}). \label{Weyl}
\end{equation}
This can actually be interpreted as the Hamiltonian for a
particle obeying the Weyl equation, by defining
\begin{equation}
P_1 = V^i_1(p_i - p_{0i}) \qquad  P_2 = V^i_2(p_i - p_{0i}) 
\qquad P_3 = V^i_3(p_i - p_{0i}) 
\end{equation}
\begin{equation}
H_{new} = H - \sigma^0 V^i_0(p_i - p_{0i}) -
\sigma^0 \omega(\vec{p_0}) 
= \vec{\sigma}\cdot\vec{P} 
\end{equation}
\begin{equation}
\omega_{new} = \omega - V^i_0(p_i - p_{0i}) - 
\omega(\vec{p_0}) 
\label{Hnew}
\end{equation}
and supposing that the $V^i_0$ are not too large\footnote{If the 
$V^i_0$ are very large, there is a risk that different sides 
of the upper light-cone fall above and below the value 
$\omega(\vec{p_0})$ of the energy at the tip of the cone.} 
compared to the other $V^i_a$'s. 
The renormalisation of the energy, eq.~(\ref{Hnew}),
is the result of transforming away a constant velocity $V_0^i$ in
D dimensions carried by all the Fermions, using the change of
co-ordinates $x^{\prime i} = x^i - tV_0^i$, and measuring the 
energy relative to $\omega(\vec{p_0})$.
Note that the ``starting momentum" $\vec{p_0}$ will generically be of the 
order of a fundamental (Planck scale) momentum, which 
cannot be significantly changed by a ``poor physicist". 
So the large momentum $\vec{p_0}$ effectively plays the role 
of a conserved charge at low energy, justifying the use 
of complex fermion fields and the existence of a Fermi 
surface.

A trivial calculation
for the Weyl equation, $H_{new} \psi = \omega_{new} \psi$, leads
to a level density with a thin neck, behaving like
\begin{equation}
 \rho \propto \omega_{new}^2
\end{equation}
According to our strong assumption about homolumo-gap effects, we
should therefore imagine that the Fermi-surface in this case would
adjust itself to be near the $\omega_{new} =0$ level. Thereby there
would then be the fewest levels near the Fermi-surface.

\subsection{The cases $n\ge 3$}

For $n$ larger than $2$ one can easily find out\footnote{HBN would 
like to thank S.~Chadha for a discussion of this $n\ge 3$ case 
many years ago.} that, in the
neighbourhood of a point where the $n$ by $n$ general Hamiltonian
matrix deviates by zero from the unit matrix, there are
generically branches of the dispersion relation for the particle
states that behave in the metallic way locally, as in the case
$n=1$. This means that the level density in such a neighborhood
has contributions like that in the $n=1$ case, varying rather
smoothly and flatly as a function of $\omega$. So these cases are
not so favourable from the Homolumo-gap point of view.

\subsection{Conclusion of the various $n$ cases for the Fermion model}

The conclusion of the just given discussion of the various
$n$-cases is that, while of course the $n=0$ case is the ``best''
case from the point of view of the homolumo-gap, it would not be
noticed by the ``poor physicist'' and thus would not be of any
relevance for him. The next ``best'' from the homolumo-gap point
of view is the case $n=2$ of just two complex components
(corresponding to 4 real components) being relevant near the
Fermi-surface. Then there is a neck in the distribution of the
levels, which is not present in the cases $n=1$ and $n>2$.

So the ``poor physicist'' should in practice observe the case
$n=2$, provided the homolumo-gap effect is sufficiently 
strong (a perhaps suspicious assumption).

Now, as we saw, this case of $n=2$ means that the Fermion field
satisfies a Weyl equation, formally looking like the Weyl equation
in just 3+1 dimensions! It should however be noticed that there
are indeed more spatial dimensions, by assumption, in our model.
In these extra spatial dimensions, the Fermions have the same
constant velocity which we were able to renormalise to zero, because
the Hamiltonian only depends on the three momentum
components $\vec{P}$ in the Taylor expandable region accessible to
the ``poor physicist''. The latter comes about because there are
only the three non-trivial Pauli matrices that make the single
particle energy vary in a linear way around the point of expansion.
In this sense the number of spatial dimensions comes out as equal to
the number of Pauli matrices.

\section{Conclusion, r\'{e}sum\'{e}, discussion}

We have found the remarkable result that, in the free
approximation, our very general quantum
field theory, which does {\em not have Lorentz invariance put in},
leads to {\em Lorentz invariance in three plus one dimensions}
for {\em both Bosons and Fermions}. In the
derivation of this result, we made use of what we called the
homolumo-gap effect and its ``analogue for Bosons'' and that
experimentalists only have access to energies low compared to the
fundamental scale. The derivation of three spatial dimensions
should be understood in the sense that our model, which has at
first a space of D dimensions, leads to a dispersion relation
(\i.e.~a relation between energy and momentum) for which the
derivative of the energy ${\omega}$ w.r.t.~the momentum in $D-3$
of the dimensions is independent of the momentum.
Then, in the remaining $3$ dimensions, we get
the well-known Lorentz invariant dispersion relations both in the
Bosonic and the Fermionic cases. In fact we obtained the Weyl
equation and the Maxwell equations, in the fermionic and the bosonic
cases respectively, as ``generic'' equations of motion -- after the
use of the homolumo gap and its analogue. These
Maxwell and spin one half equations of motion are in remarkable
accord with the presently observed (\i.e. ignoring the Higgs particle)
fundamental particles!

\subsection{Some bad points and hopes}
In spite of this remarkable success of our model in the free
approximation, we have to admit there are a number of flaws:

1) The three space dimensions selected by each type of particle
{\em are a priori overwhelming likely {\em not} to be the same
three}. That is to say we would have to hope for some
speculative mechanism that could align the three dimensions used
by the different species of particles, so as to be the same three
dimensions.

2) Although we have hopes of introducing interactions, 
it is not at all clear how these 
interactions would come to look and whether e.g.~they would 
also be Lorentz invariant---according to point 1) one would
{\it a priori}
say that they do not have much chance to be Lorentz invariant.

3) There are extra dimensions in the model, although they do not
participate in the derived Lorentz invariance which is only a 3+1
Lorentz invariance. Rather the velocity components in the extra 
dimension directions are  constant, independent of the momentum 
of the particles. We can really by convention renormalise them 
to zero and claim that we do not see the extra dimensions, 
because we cannot move in these directions. 
But from point 1) there is the worry
that these directions (in which we have no movement)
are different for the different types of particles.

The best hope for rescuing the model from these problems might be to
get rid of momentum conservation in the extra directions. We might
hope to get some attachment of the particles to a fixed position
in the extra directions much like attachments to branes, but then
one would ask how this could happen in a natural way.
Of course the point of view most in the spirit of the random 
dynamics project would be that {\it a priori} we did not even 
have momentum conservation, but that
it also just arose as the result of some Taylor expansion. This
becomes very speculative but it could easily happen that it is
much easier to get a translational invariance symmetry develop,
along the lines suggested in section 6.2.3 of our book \cite {book}, 
for the momentum directions in which we have rapid motion than 
in the directions in which we have zero velocity. If we crudely
approximated the particles by non-relativistic ones, the rapid
motion would mean low mass while the zero motion
would mean a very huge mass. The uncertainty principle would, 
therefore, much more easily allow these
particles to fall into the roughness valleys\footnote{By
roughness valleys we refer to the (local) minima or valleys
in a potential representing a non-translational invariant
potential set up so as break momentum conservation.} in the translational
invariance violating potential in the extra directions, where the
non-relativistic mass is much larger than in the 3 space directions.
A particle would be very much spread out by uncertainty in the
3 directions and, thus, only feel a very smoothed out roughness
potential, if translational invariance is broken in these
directions. In this way translational invariance could develop
in just 3 dimensions.

A breakdown of the translational invariance---or, as just
suggested, a lack of its development---in the extra
dimensions would be very helpful in solving the  above-mentioned
problems. This is because there would then effectively only be 3
space dimensions and all the different types of particles would,
thus, use the same set of 3 dimensions. It must though be admitted that
they would still have different metric tensors, or metrics we
should just say. We had some old ideas \cite{book} for solving
this problem, but they do not quite work in realistic models.

\subsection{Where did the number three for the space dimension come from?}

One might well ask why we got the prediction of just three
for the number of spatial dimensions. In fact we
have derived it differently, although in many ways analogously,
for Bosons and for Fermions:

\noindent {\bf Bosons:} \  For Bosons we obtained this result by
considering the simultaneous approach of a pair of equal and
opposite eigenvalues of the real antisymmetric matrix $A_{ij}$ to
the supposedly existing zero frequency, $\omega = 0$, level(s).
Thus the rank of the matrix $A$ relevant to this low energy range
is just {\em two}, except at the point around which we expand
where it has rank zero. Then we argued that we could transform
such a matrix in such a way that it effectively becomes a 3 by 3
matrix---still antisymmetric and real. So the matrix $A$ has
effectively three independent matrix elements and each can vary
with the components of the momentum. However, in the low energy
regime, this dependence can be linearized and $A$ only depends on
three linearly independent components of the momentum. It is these
three dimensions in the directions of which we have non-zero
velocity (or better non-constant velocity) for the Boson---the
photon, say, in as far as it obeys the Maxwell equations. We thus
got the number three as the number of independent matrix elements
in the antisymmetric matrix $A$, obtained after transforming away
most of this rank two matrix.

\noindent {\bf Fermions:} \  In this case we went to a complex
notation, although we still started from the same type of
antisymmetric real matrix as in the bosonic case. The homolumo-gap
argumentation suggested that just $n=2$ complex components in the
field should be ``relevant'' near the Fermi surface, after ruling
out the trivial $n=0$ case as unobservable by anybody. This number
of relevant components then meant that just $n^2 -1=4-1=3$
non-trivial linearly independent $n$ by $n$ matrices could be
formed. These three matrices could, of course, then be used as the
coefficients for three momentum components in the linearized
(Taylor expanded) momentum dependence of the Hamiltonian. In this
way the number three arose again.

So there is an analogy at least in as far as, for both Bosons
and Fermions, it is the number of linearly independent matrices of
the type finally used, which remarkably predicts the observable
number of spatial dimensions to be 3. However in the bosonic case
it is real three by three matrices which we ended up with, while
in the fermionic case it is Hermitian 2 by 2 matrices with the
unit matrix omitted. The unit matrix is not counted because it
does not split the levels and basically could be transformed away
by the shift of a vierbein, \i.e.~by adjusting the meaning of the
momentum and energy components.

A strange prediction of the Boson model is that, at first at
least, we get a parity symmetric state of the world for the
Maxwell fields. That is to say for every state of the
electromagnetic---or generalized Yang Mills---field there is
somewhere, reflected in the origin of position space, a
corresponding reflected state. In principle we could test such an
idea, by looking to see whether we could classify galaxies found on
the sky into pairs that could correspond to mirror images---in the
``origin''---of each other. Really we hope that this illness of
our model might easily repair itself.

\section{Acknowledgements}
C.D.F.~thanks Jon Chkareuli and H.B.N.~thanks Bo Sture Skagerstam 
for helpful discussions. C.D.F.~thanks PPARC for a travel grant to 
attend the 2001 Bled workshop and H.B.N.~thanks the Alexander von 
Humboldt-Stiftung for the Forschungspreis.
 


\begin{thebibliography}{99}


\bibitem{RDold}
H.~B.~Nielsen, {\it Fundamentals of Quark Models}, eds. I.~M.~Barbour
and A.~T.~Davies, Scottish Universities Summer School in Physics
(1976), pp. 528-543.

\bibitem{book}
C.~D.~Froggatt and H.~B.~Nielsen, {\it Origin of Symmetries},
(World Scientific, Singapore, 1991).

\bibitem{NormaHolger}
N.~Mankoc Borstnik and H.~B.~Nielsen, hep-ph/0108269.

\bibitem{RughHolger}
H.~B.~Nielsen and S.~E.~Rugh, hep-th/9407011.

\bibitem{Kostelecy}
{\em CPT and Lorentz Symmetry}, 
ed.~by V.~A.~Kostelecky
(World Scientific, Singapore, 1999);\\
{\em CPT and Lorentz Symmetry II}, 
ed.~by V.~A.~Kostelecky
(World Scientific, Singapore, 2002) 

\bibitem{Teller}
H.~A.~Jahn and E.~Teller, Proc. Roy. Soc. London
{\bf A161} (1937) 220.

\bibitem{Colladay}
D.~Colladay and V.~A.~Kostelecky, 
Phys. Rev. {\bf D55} (1997) 6760;\\
D.~Colladay and V.~A.~Kostelecky, 
Phys. Rev. {\bf D58} (1998) 116002;\\
D.~Colladay and P.~McDonald, 
J.~Math.~Phys.~{\bf 43} (2002) 3554.

\end{thebibliography}
\end{document}